\documentclass[reprint,amsmath,amssymb,aps,pra]{revtex4-1}
\usepackage[utf8]{inputenc}
\usepackage{graphicx}
\usepackage{dcolumn}
\usepackage{mathrsfs,amsmath}
\usepackage{bm}
\def\ph2{{\it p}-H$_2$}
\def\Am3{\AA$^{-3}$}

\renewcommand{\vec}[1]{\mathbf{#1}}
\begin{document}
\title{Dynamic structure factor of superfluid $^4$He from  
\\ Quantum Monte Carlo: Maximum Entropy revisited}
\author{Youssef Kora and Massimo Boninsegni}
\affiliation{Department of Physics, University of Alberta, Edmonton, Alberta, T6G 2E1, Canada}
\date{\today}

\begin{abstract}
We use the Maximum Entropy Method (MaxEnt) to estimate the dynamic structure factor of superfluid $^4$He at $T=1$ K, by inverting imaginary-time density correlation functions computed by Quantum Monte Carlo (QMC) simulation. Our procedure consists  of a Metropolis random walk in the space of all possible spectral images, sampled from a probability density which includes the entropic prior, in the context of the so-called ``classic" MaxEnt. Comparison with recent work by other authors shows that, contrary to what is often stated, sharp features in the reconstructed image are not ``washed out" by the entropic prior if the underlying QMC data have sufficient precision. Only {\em spurious} features that tend to appear in a straightforward $\chi^2$ minimization are suppressed.
\end{abstract}
\maketitle

\section{Introduction}
Quantum Monte Carlo simulations are among the most reliable tools to investigate the physics of quantum many-body systems in thermal equilibrium. In particular, thermodynamic properties of interacting Bose assemblies, such as superfluid $^4$He, can be calculated quite accurately \cite{worm}.
At least in principle, QMC also allows one to obtain dynamical properties, at least within the linear response approximation; for, one can compute correlation functions in imaginary time, from which spectral functions can be inferred through an inverse Laplace transformation.    
\\ \indent 
Unfortunately, the inversion is mathematically ill-posed, and because QMC data are inevitably affected by statistical uncertainties, an unambiguous determination of the spectral function is usually not possible. In some cases, prior knowledge about the physics of the system may constrain the set of possible solutions, allowing for a reliable reconstruction; for example, one may know that the spectral function is dominated by one or two well-defined peaks, and simply fit the QMC data accordingly (see, for instance, Ref. \onlinecite {saccani}). \\ \indent 
In the general case, however, when no such knowledge is available, a large number of very different images will be consistent with the QMC data. Thus, one will typically resort to some kind of ``regularization'' scheme (RS), aimed at retaining only those images whose non-trivial structure is {\em truly warranted by the data}. Consequently, any RS will inevitably tend to soften some of the sharpest features; for example, distinct, isolated peaks will be broadened, to reflect the inherent uncertainty arising from the finite precision of the data and the ill-posedness of the problem \cite{justsoyouknow}. 
\\ \indent 
A popular RS, in the context of inversion of QMC data, is the Maximum Entropy method (MaxEnt) \cite{probstat,bayesian}, which has been applied to the determination of spectral functions of various lattice many-body Hamiltonians \cite{gubernatis,white,makivic,bulut,preuss} as well as of the dynamic structure factor in normal and superfluid $^4$He \cite{massimo1996}. In general, MaxEnt has yielded quantitatively reliable results for some of the main aspects of the reconstructed images, i.e., the positions of the peaks, and therefore the determination of the excitation spectrum; on the other hand, the quantitative accuracy of predictions concerning, e.g., the widths of the peaks, and the ensuing ability to resolve adjacent peaks, was less satisfactory, although in most cases the limiting factor was the quality of the QMC data, rather than the RS adopted to extract the images. 
Alternative RS have been proposed in the course of the years,  the context of QMC simulations \cite{sandvik,sylvjuasen,mishenko,reichman,fuchs}, displaying some advantages over others for specific applications, but no comprehensive, systematic comparison has yet  been carried out (at least to our knowledge). 
\\ \indent 
In recent years, the problem of extraction of the dynamic structure factor of superfluid $^4$He from imaginary-time correlations computed by QMC has been independently revisited by two groups \cite{gift,ferre}, who proposed RSs not making use of MaxEnt's entropic prior. In both cases, their procedure essentially amounts to $\chi^2$-fitting \cite {diff},
supplemented by averaging over a set of comparable images, in order to suppress some of the spurious structure that inevitably arises on carrying out $\chi^2$ minimization in the presence of an ill-posed problem. Both works make the claim that their proposed approaches are superior to MaxEnt, in that the resulting images are sharper and in better agreement with experimental data.
\\ \indent 
In this paper, we revisit the use of MaxEnt for the same problem, in order to assess quantitatively the claims made in Refs. \onlinecite {gift,ferre}. Specifically, we estimate the dynamic structure factor $S({\bf q},\omega)$ for superfluid $^4$He, by computing imaginary-time density correlations by QMC, and by using MaxEnt  to carry out the inversion. Our methodology is similar to that of Ref. \onlinecite{massimo1996}, i.e., it consists of a Metropolis random walk in the space of spectral images, sampled from a probability density proportional to the standard maximum likelihood estimator, multiplied by the entropic prior (see below). This procedure allows us to assign an uncertainty in the value of $S({\bf q},\omega)$, as the standard deviation of the values recorded for the different frequencies in the course of the random walk.
\\ \indent 
Compared to Ref. \onlinecite{massimo1996}, our present study obviously benefits from two decades of advances, both in computing hardware as well as in the QMC methodology utilized to generate the imaginary-time data. As a result, our statistical uncertainties are much smaller than those of the 1996 work, comparable to those of the data used in Refs. \onlinecite{gift,ferre}, which is a necessary condition in order to carry out a meaningful and fair comparison. 
Based on the results presented here, we contend that MaxEnt does {\em not} prevent sharp features from appearing in the reconstructed spectral functions, as long as the accuracy of the QMC data justifies their inclusion. Indeed, the spectral images shown here are of comparable (or better) quality than those offered in Refs. \onlinecite{gift,ferre}. Ultimately, the sharpness of the spectral image almost exclusively hinges on the accuracy of the QMC data; by promoting smoothness, the entropic prior serves in our view a useful, noise-reducing purpose. 
\\ \indent
It is worth noting that a general scheme capable of tackling this kind of problem can be applied in other, rather different contexts, e.g., the determination of ground state expectation values in QMC transient estimate calculations \cite{caffarel}. These are typically carried out for Fermi systems, which are affected by the infamous ``sign'' problem, resulting in an exponential increase with imaginary time of the statistical error (see, for instance, Ref. \onlinecite{bm}).
\\ \indent 
The remainder of this paper is organized as follows: in section \ref{mm} we describe the model of the system and the QMC calculations carried out in this work; in Sec. \ref{me} we describe in detail our inversion method; we present and discuss our results in Sec. \ref{res} and finally outline our conclusions in Sec. \ref{conc}.

\section{Model and QMC calculation}\label{mm}
In this section we describe the QMC calculation of the imaginary-time correlation function which is then inverted to obtain the dynamic structure factor. 
The system is described as an ensemble of $N$ point-like, identical particles with  mass $m$ equal to that of a He atom and with spin $S=0$, thus  obeying Bose statistics.  It is enclosed in a cubic cell, with periodic boundary conditions in the three directions. 
The quantum-mechanical many-body Hamiltonian reads as follows:
\begin{eqnarray}\label{u}
\hat H = - \lambda \sum_{i}\nabla^2_{i}+\sum_{i<j}v(r_{ij})
\end{eqnarray}
where the first (second) sum runs over all particles (pairs of particles), $\lambda\equiv\hbar^2/2m=6.0596415$ K\AA$^{2}$, $r_{ij}\equiv |{\bf r}_i-{\bf r}_j|$ and $v(r)$ is a pair potential which describes the interaction between two atoms. We make use in this study of the accepted Aziz pair potential \cite{aziz79}, which has been utilized in most simulation studies of  superfluid helium. A more accurate model would also include interactions among triplets of atoms; however, published
numerical work has given strong indications
that three-body corrections, while significantly affecting the
estimation of the pressure, have a relatively small effect on
the structure and dynamics of the system, of interest here \cite{mpfb}.
\\ \indent 
We carried out QMC simulations of the system described by Eq.  (\ref{u}) at temperature $T = 1$ K, using the continuous-space Worm Algorithm \cite{worm}.  Since this technique is by now fairly well-established, and extensively described in the literature, we shall not review it here. A canonical variant of the algorithm was utilized, in which the total number of particles $N$ is held fixed \cite{mezz1,mezz2}. 
\\ \indent 
The quantity of interest here is the dynamic structure factor $S({\bf q},\omega)$, which describes density fluctuations of wave vector {\bf q}. For superfluid $^4$He it has been extensively studied experimentally by neutron scattering (for a review, see, for instance, Ref. \onlinecite{glyde}). It is a direct probe of the elementary excitations (phonons and rotons) that underlie the physical behavior of the system at low temperature \cite{landau,feynman,cohen}.  
$S({\bf q},\omega)$ is non-negative function satisfying the
relation \cite{lovesey}
\begin{equation}\label{fsum}
\langle\omega\rangle =\int_0^\infty\ d\omega\ \omega\ S({\bf q},\omega)\ (1-e^{-\beta\omega})=\frac{{q}^2}{2m}
\end{equation}
known as {\em f-}sum rule (we henceforth set $\hbar=1$, the Boltzmann constant $k_B=1$ and define $\beta=1/T$). 
There is no known QMC scheme allowing for the {\em direct} calculation of $S({\bf q},\omega)$. However, it can be shown (see, for instance, Ref. \onlinecite{massimo1996}) that
\begin{align}\label{fourier2}
F(\vec{q},\tau) = \int_0^{\infty}\ d\omega\ (e^{-\omega \tau}+
e^{-\omega(\beta-\tau)})\ S(\vec{q},\omega)
\end{align}
where $0\le\tau\le\beta$ and
$F({\bf q},\tau)$ is the imaginary-time auto-correlation function 
\begin{equation}\label{propagator}
F(\vec{q},\tau)=\frac{1}{N}\ \langle\hat\rho_{\vec{q}}(\tau)\ \hat\rho_{\vec{q}}^\dagger(0)\rangle 
\end{equation}
where $\langle ...\rangle$ stands for thermal average, and with
\begin{align}\label{densityq}
\rho_{\vec{q}}({\tau}) = \sum_{j=1}^N  e^{i \vec{q} \cdot \vec{r}_{j}},
\end{align}
where the $\{{\bf r}_j\}$, $j=1,2,...N$ are the positions of the $N$ $^4$He atoms at imaginary time $\tau$ along the many-particle path. 
The quantity $F({\bf q},\tau)$ is what is actually computed by QMC, for a discrete set of values of $\tau$; $S({\bf q},\omega)$ is inferred from $F({\bf q},\tau)$ through a numerical inversion of eq. \ref{fourier2}. The details of this procedure are outlined in Sec. \ref{me}.
\\ \indent 
The QMC simulation is  standard;  we adopted the usual the short-time approximation to the imaginary-time propagator accurate to fourth order in the time step $\epsilon$ (see, for instance, Ref. \onlinecite{jltp}).  All of the results presented here are extrapolated to the $\epsilon\to 0$ limit; just like for other observables, the numerical estimates of the quantities of interest here, namely the imaginary-time correlation functions described below, computed with a value of the time step $\epsilon=$ (1/640) K$^{-1}$ are indistinguishable from the extrapolated ones, within the statistical uncertainties of the calculation.
\\ \indent 
Calculations were carried out at two different densities, namely 0.021834 \Am3, which is that at saturated vapor pressure (SVP) \cite{barenghi}, and 0.0260 \Am3, which is very close to the freezing density (at a pressure of approximately 25 bars). All calculations were carried out at $T=1$ K. The experimental and theoretical data we compare our results against are at temperatures that range from 0 K to 1.3 K. All such temperatures are well below the lambda transition, and at that level the excitations are essentially independent of temperature (see, for instance, Refs. \onlinecite{gibbs,dietrich1972}).
We took advantage of space and time symmetry to improve statistics; a rough estimate of the statistical error on the generic value of $F({\bf q},\tau)$ is given by $5\times 10^{-4}\ F({\bf q},0)$.
\\ \indent 
The bulk of the results shown here were obtained on a system comprising $N=64$ particles, a number which is not particularly large but that allows us to collect good statistics in a given simulation time; experience with previous work \cite{massimo1996} suggests that this system size is sufficient to extract information at the wave vectors of interest here (see below). However, we have also repeated the simulation with $N=256$ particles, and found no statistically significant difference in the values of $F({\bf q},\tau)$, within the statistical errors of our calculation. 
\\ \indent 
$F({\bf q},0)\equiv S_{\bf q}$ is known as the {\em static structure factor}, which is experimentally accessible and it is related via a Fourier transformation to the atomic pair correlation function. The values of  $S_{\bf q}$ obtaind here are in quantitative agreement with previous calculations, i.e., in excellent agreement with experiment (see Ref. \onlinecite{rmp}). 
\begin{figure}[h]
\centering
\includegraphics[width=0.47\textwidth]{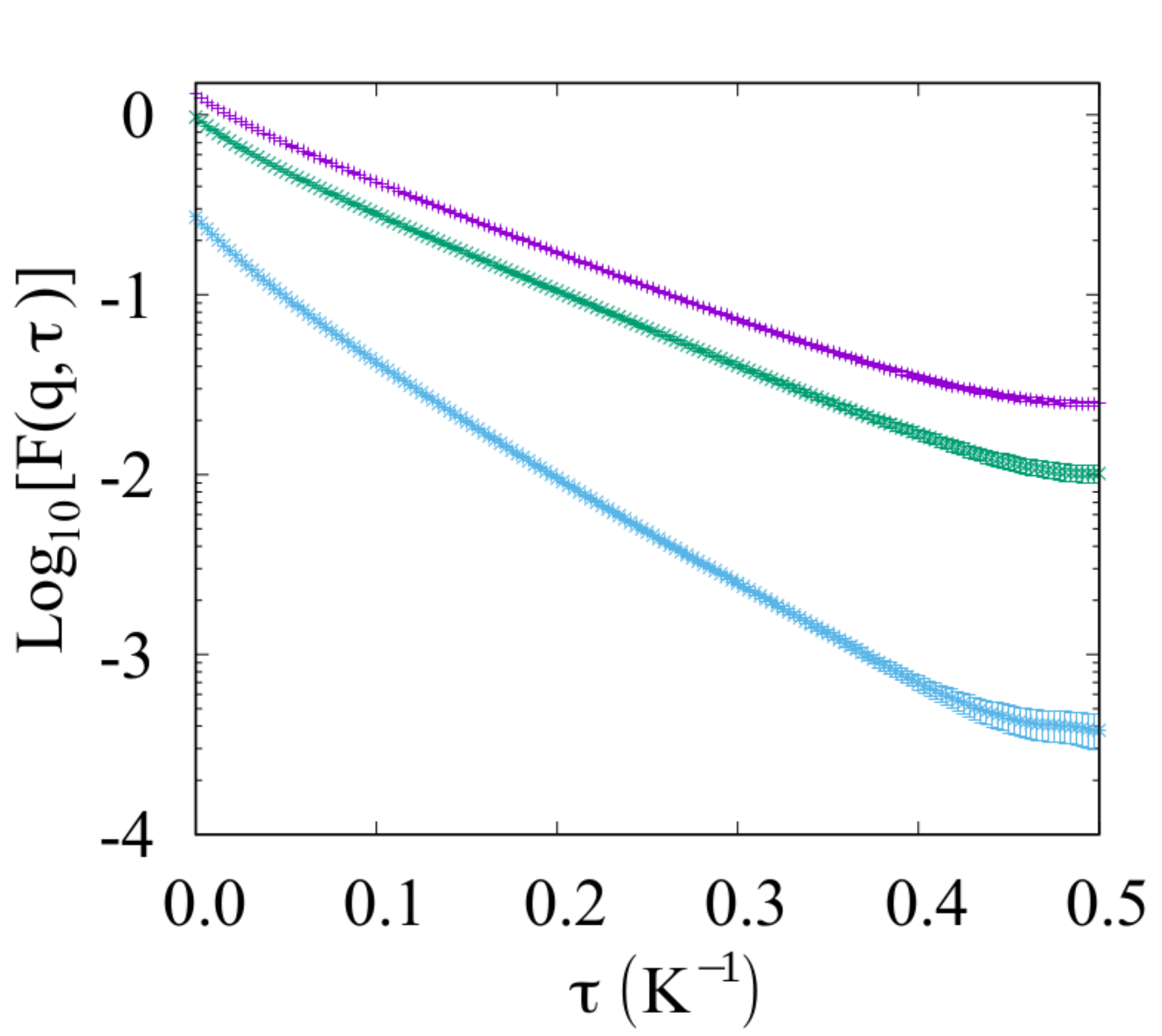}
\caption{{\em Color online}. Typical $F(\vec{q},\tau)$ results computed in  a simulation of superfluid $^4$He at $T=1$ K at density 0.021834 \AA$^{-3}$. Results shown here are for the wave vectors $q=1.075$ \AA$^{-1}$ (bottom curve), $q=1.756$ \AA$^{-1}$ (middle curve) and $q=1.964$ \AA$^{-1}$ (top curve). When not shown, statistical errors are smaller than the size of the symbols.}
\label{f1}
\end{figure}
\\ \indent
Typical results for $F({\bf q},\tau)$ are shown in Fig. \ref{f1}; because $F({\bf q},\tau)=F({\bf q},\beta-\tau)$ (see, for instance, Ref. \onlinecite{lovesey}), one need only compute this quantity in the $0\le\tau\le\beta/2$ interval.
\section{MaxEnt Inversion}\label{me}
The problem with the numerical inversion of eq. \ref{fourier2}, aimed at obtaining $S({\bf q},\omega)$ from the values of $F({\bf q},\tau)$ computed by QMC, lies in the fact that the integral kernel exponentially suppresses the contribution at high frequency of the spectral function to $F({\bf q},\tau)$; consequently, $F(\vec{q},\tau)$ is minimally affected by the high frequency behavior of $S({\bf q},\omega)$. Because $F(\vec{q},\tau)$ is the result of QMC simulations, and therefore possesses finite statistical uncertainties, there will be typically a large set of physically different spectral functions consistent with the numerical data for $F(\vec{q},\tau)$. Most of these solutions are unphysical and/or bear little resemblance to the actual  $S({\bf q},\omega)$. The goal is that of finding a systematic and robust way to weed out spurious solutions, and retaining only a relatively small subset of physical ones, from which at least the most important physical features of $S({\bf q},\omega)$ may be reliably extracted.
\\ \indent 
As mentioned above, $F({\bf q},\tau)$ is computed for the discrete  set of imaginary times $l\delta\tau$, $l=0,1,...,L$, with $2L\delta\tau=\beta$.
In order to simplify the notation, for a given value of {\bf q} we define $\vec{F}\equiv\{F_0,...,F_L\}$, with $F_l\equiv F({\bf q},l\delta\tau)$. Each entry $F_l$ is affected by a statistical uncertainty $\sigma_l$, estimated by careful binning analyses of data (see, for instance, Ref. \onlinecite{flyvbjerg1989}) collected over sufficiently long simulations. 
We begin by approximating the integral on the right hand side of eq. \ref{fourier2} with a sum, i.e., turn eq. \ref{fourier2} into a system of algebraic equations that can be expressed in compact matrix form
\begin{equation}\label{mainrelation3}
\vec{F} = \vec{K} \vec{S},\end{equation} 
having defined
\begin{align}\label{Kmat}
K_{lj} =  [e^{-jl \delta \omega \delta \tau} + e^{-j(2L-l)\delta \omega \delta\tau}]\ \delta \omega,
\end{align}
$\vec{S} \equiv \{ S_1,...,S_M \}$, $S_j\equiv S({\bf q}, j\delta\omega)$, and $M\delta\omega=\omega_M$, $\omega_M$ chosen large enough that $S({\bf q},\omega)$ can be set to zero for $\omega > \omega_M$, and $\delta\omega$ small enough to achieve the desired frequency resolution. In this study, $\omega_M$ is between 100 and 300 K, whereas $M$ is between 150 and 400. An important observation is that typically $M>L$, i.e., the system (\ref{mainrelation3}) is underdetermined, and 
therefore, in general, no unique solution can be found, quite irrespective of the ill-posedness of the problem and of statistical errors of the computed imaginary-time correlation functions \cite{ts}. 
\\ \indent 
We take the same approach as in Ref. \onlinecite{massimo1996}, based on ``classic'' MaxEnt (see, for instance, Ref. \onlinecite{bayesian}) and define our ``optimal'' solution as
\begin{equation}\label{optimal}
{\vec S}_{\circ} \equiv \int d\alpha\ \int \mathscr{D}{\vec S}\ {\vec S}\ {\cal F}(\alpha,{\vec S})
\end{equation}
where $\mathscr{D}{\vec S}\equiv dS_1dS_2...dS_M$, and
\begin{equation}\label{sample}
{\cal F}({\alpha,\vec S})=\frac{e^{-\chi^2({\vec S})/2}}{Z_Q}\ \frac{e^{\alpha {\cal S}({\vec S})}}{Z_{\cal S}(\alpha)}\ \rho({\vec S})
\end{equation}
is a prior probability assigned to the generic image $\vec S$.
Here, $\alpha$ is a non-negative regularization parameter, to which we come back below; $Z_Q$ and $Z_S(\alpha)\propto \alpha^{-M/2}$ are normalization constants;
\begin{equation}\label{chisq}
\chi^2({\vec {\bar S}}) = (\vec{F}-\vec{\bar{F}})^T\vec{C}^{-1} (\vec{F}-\vec{\bar{F}})
\end{equation}
is the standard measure of goodness of fit, with ${\vec{\bar F}}={\vec K}{\vec {\bar S}}$ and we make the diagonal approximation \cite{noteco} for the covariance matrix $\vec{C}$, i.e., 
\begin{equation}\label{covariance}
C_{ij}=\sigma^2_{i}\delta_{ij},
\end{equation}
and
\begin{equation}\label{jay}
{\cal S}(\vec{S}) =-\sum_{i=1}^M \ f_i\ ln\biggl ( {Mf_i}\biggr ),
\end{equation}
with $f_i=S_i/(\sum_j S_j)$, is Jaynes' entropy of the image ${\vec S}$ \cite{jaynes,zipo}; and finally,
\begin{equation}\label{fsumprob}
\rho(\vec{S}) \propto {\rm exp} \left(-\frac{[\langle\omega\rangle-\omega_\vec{q}]^2}{2\eta^2\omega^2_\vec{q}}\right)
\end{equation}
where $\langle\omega\rangle$ is defined in eq. \ref{fsum}, $\omega_{\bf q}=q^2/(2m)$ and $\eta$ is adjusted to enforce that relation (\ref{fsum}) be satisfied to the desired degree of accuracy (typically $\eta\le0.01$).
\\ \indent
The prior probability (\ref{sample}) ascribes greater weight to those spectral functions that are consistent with the data, and therefore have a low value of $\chi^2$ and fulfill the $f$-sum rule, while at the same time are smoother in character. In other words, sharp features such as isolated peaks should {\em not} be included unless consistency with the data requires it.
\\ \indent 
The parameter $\alpha$ can be used to ``tune'' the relative importance of the entropic prior in ${\cal F}(\vec{S})$; in the limit $\alpha\to 0$, one is performing conventional $\chi^2$-fitting; on the other hand, as $\alpha$ grows the entropic prior becomes increasingly important. The question arises of how to choose the value of $\alpha$. In ``historic'' MaxEnt, one adjusts $\alpha$ so that on average, the value of $\chi^2\sim L$. As mentioned above, we adopt the ``classic'' MaxEnt approach, in which $\alpha$ is treated as a random variable, and assigned a prior probability distribution $p(\alpha)$, which is incorporated in the normalization constant $Z_S(\alpha)$. 
\\ \indent 
We evaluate the multidimensional integral in eq. \ref{optimal} by Monte Carlo, just as in Ref. \onlinecite {massimo1996}. Specifically, we perform 
a random walk in  $\{ \vec{S},\alpha\}$-space, using the Metropolis algorithm to sample the 
probability density given by eq. \ref{sample}.
We achieve that through few elementary moves, designed to satisfy the usual detailed balance condition. Specifically, we randomly attempt either one of the following:
\begin{enumerate}
\item{ the displacement of an elementary amount of area, equal to $\gamma\ \delta S$, where $0\le \gamma\le 1$ is a uniform random number, from a randomly selected channel $j$ to another one, randomly selected among $j-p,...j-1,j+1,... j+p$.}
\item
{the addition or subtraction of $\gamma\ \delta S^\prime$ from a randomly selected channel $j$. }
\item{the change of $\alpha$ by an amount $(1/2-\gamma) \ \delta\alpha$.}
\end{enumerate}
Proposed moves are accepted or rejected based on the usual Metropolis test, making use of eq. \ref{sample} in the acceptance ratio  \cite{noneg}.
The parameters $\delta S$, $\delta S^\prime, \delta\alpha$ and $p$ are adjusted to ensure a 50\% acceptance rate. The move attempting to change the value of $\alpha$ is typically attempted every $\sim M$ attempts to perform either one of the first two moves. 
\begin{figure}[h]
\centering
\includegraphics[width=0.47\textwidth]{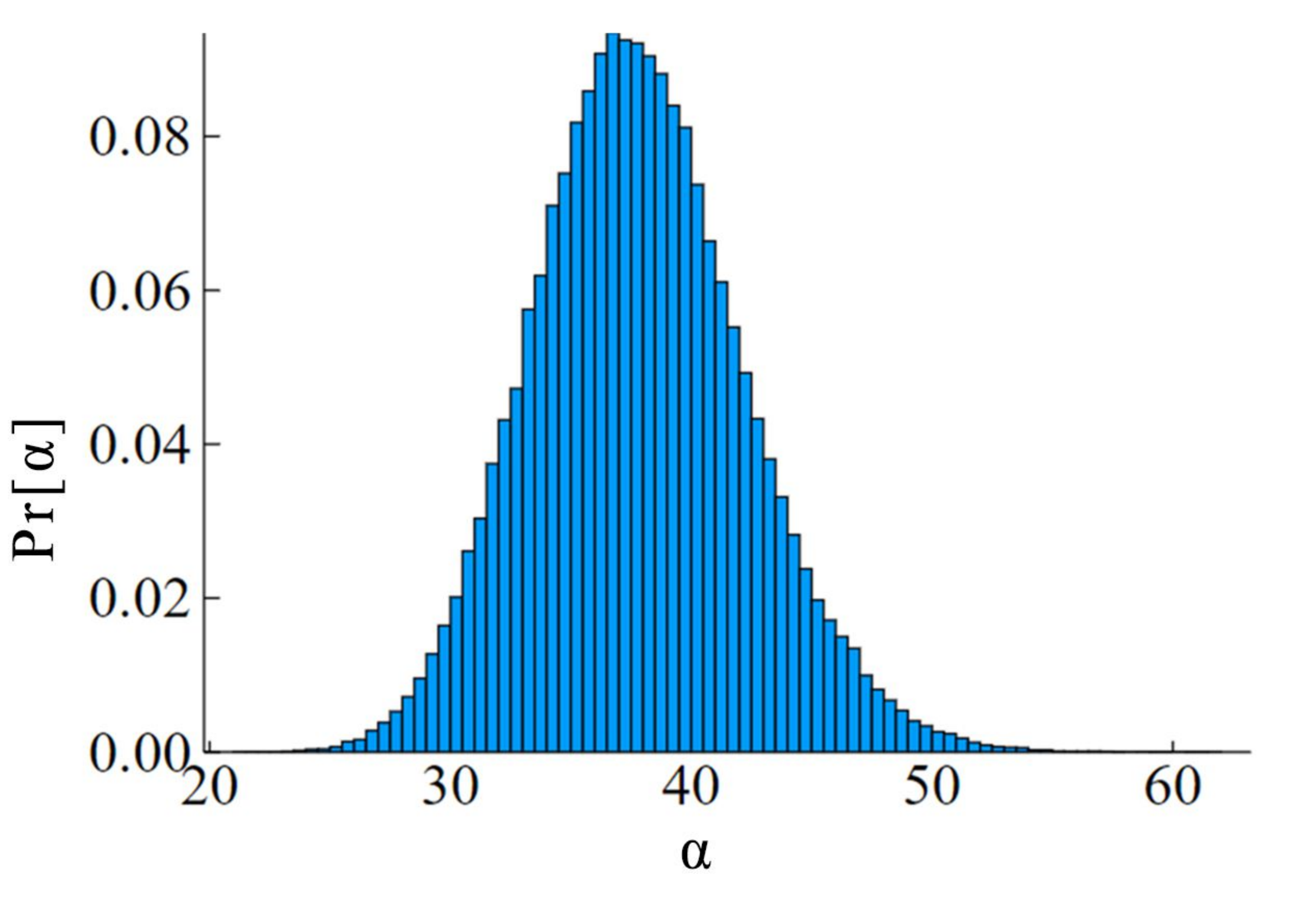}
\includegraphics[width=0.47\textwidth]{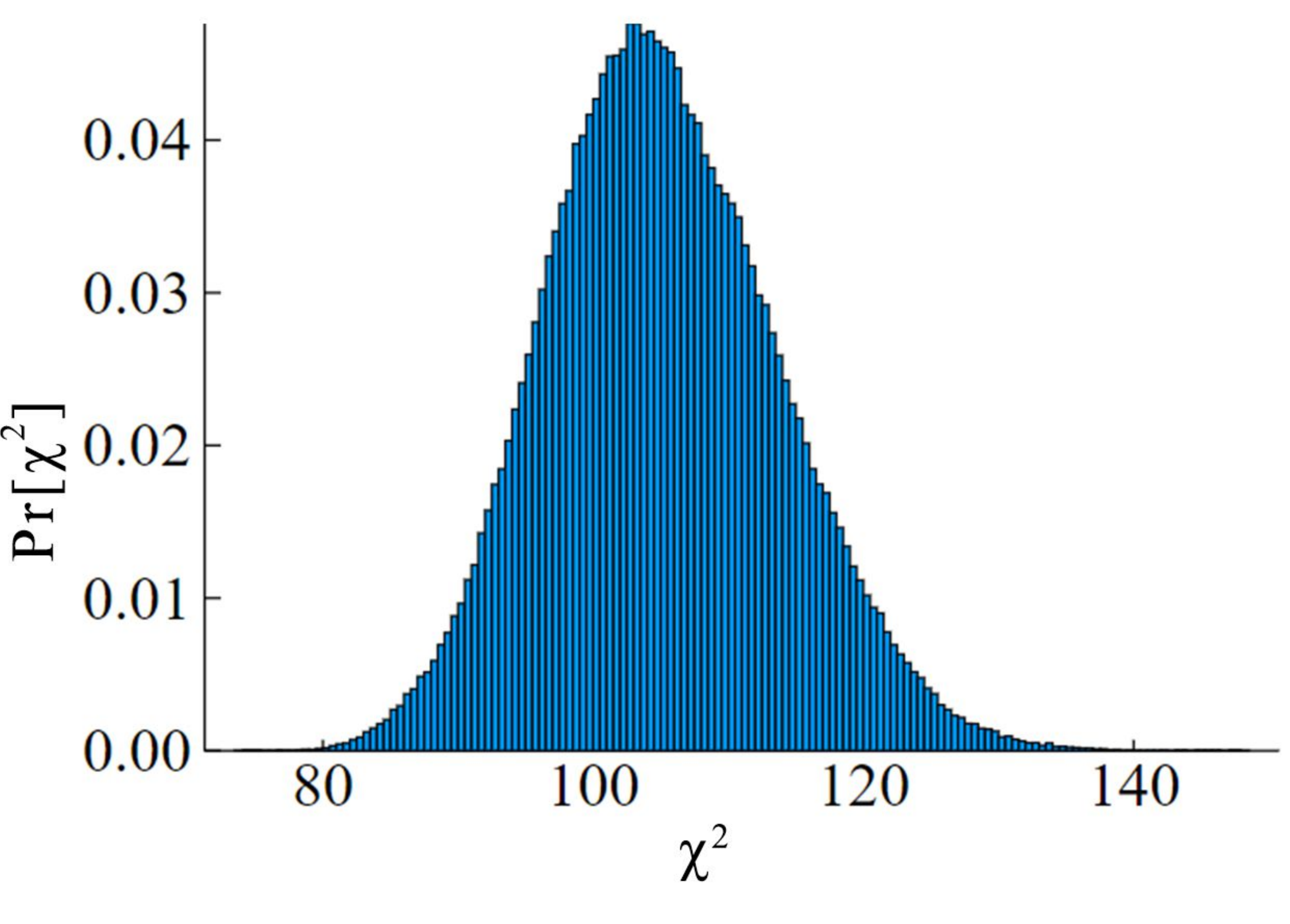}
\caption{Posterior probability for the regularization parameter $\alpha$ (top) and for the
the goodness-of-fit parameter $\chi^2$ (bottom), obtained from the Metropolis random walk in $\{\vec S,\alpha\}$-space as described in the text. This particular result refers to the $q=1.756$ \AA$^{-1}$ case.  }\label{f1b1}
\end{figure}

The posterior probability of $\alpha$, $Pr[\alpha]$ as well as the $\chi^2$ distribution $Pr[\chi^2]$, are obtained from the random walk, just as in Ref. \onlinecite{massimo1996}. Fig. \ref{f1b1} shows a typical result.
\\ \indent 
The optimal image $\vec S_\circ$ (eq. \ref{optimal}), determined as an average over the images generated in the random walk, is affected by a statistical error, that can be estimated in the standard way, and can be rendered sufficiently small upon using a relatively modest amount of CPU time. More significant, however, given the inherent uncertainty of the inversion, is the {\em standard deviation} associated with the fluctuation of the values $S_i$ around their averages; we report it below, when illustrating our results, as it furnishes in our view a fair assessment of the range of variation of the solution.

\section{Results}\label{res}

\begin{figure}[h]
\centering
\includegraphics[width=0.47\textwidth]{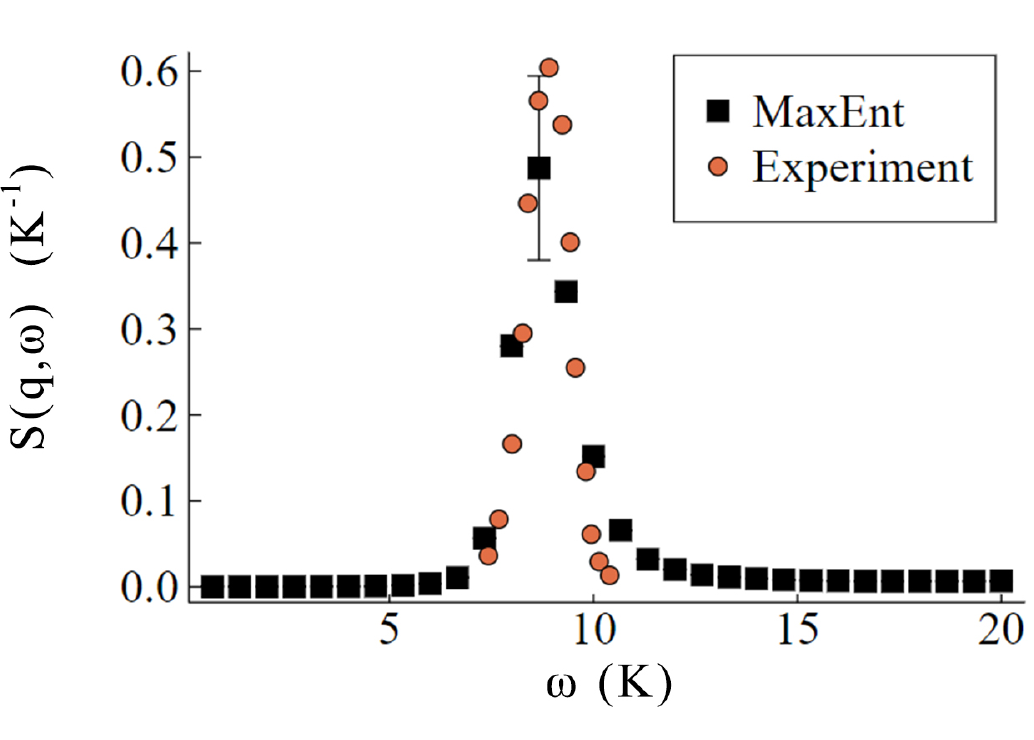}
\caption{{\em Color online}. $S(\vec{q},\omega)$ in superfluid $^4$He at $T=1$ K (at SVP) for the roton wave vector ($q=1.963$  \AA$^{-1}$), computed by inversion of QMC data based on MaxEnt (eq. \ref{optimal}, squares).  Statistical errors on $S(\vec{q},\omega)$ are smaller than the sizes of the symbols; the error bar on the square represents the {\em standard deviation} (see text), which has similar values for the two data points adjacent to the peak, and is comparable to, or smaller than symbol sizes for all other data points. Circles show experimental data from Ref. \onlinecite{andersen1994} (only the coherent part is shown) at $T=1.3$  K for the wave vector $q=1.90$  \AA$^{-1}$.}
\label{rotexp}     
\end{figure}
Fig. \ref{rotexp} shows results for $S({\bf q},\omega)$ 
for the roton wave vector ($q=1.963$ \AA$^{-1}$) 
at $T=1$ K and at saturated vapor pressure (SVP). Squares represent the values of $\vec S_\circ$  defined through eq. \ref{optimal}, computed by means of the Monte Carlo Metropolis procedure described in Sec. \ref{me}. The statistical errors on the values of $\vec S_\circ$ are smaller than the sizes of the symbols. Also shown in the figure are experimental data \cite{coherent} from Ref. \onlinecite{andersen1994} at $T=1.3$ K and for the wave vector $q=1.90$ \AA$^{-1}$. 
Agreement between theory and experiment seems fairly good; not only the position, but also the width of the peak is rather well reproduced, unlike in previous applications of MaxEnt \cite{massimo1996}. 
This result shows that MaxEnt does not prevent the reconstructed spectral image from developing sharp features, if the quality of the underlying QMC data justifies their inclusion. In the presence of greater statistical uncertainties, on the other hand, MaxEnt implies a more conservative choice, namely one in which smoother images are privileged.
\\ \indent 
As mentioned above, the statistical errors on $\vec S_\circ$ are comparable to, or smaller than the sizes of the symbols, and can always be rendered negligible with modest computing resources. Obviously, however, the issue arises of assessing {\em systematic} errors, which are inherent to this image reconstruction problem. In other words, how far off can the optimal image $\vec S_\circ$ be expected to be from the  actual spectral function? The Metropolis procedure adopted here allows us to offer an estimate of that through the standard deviation of the values of $\vec S_\circ$ for each and every value of the energy. In Fig. \ref{rotexp} we show one such standard deviation, corresponding to the energy  interval $\omega_m$ in which $\vec S_\circ$ takes on its highest value. Although not shown in the figure for clarity, $\vec S_\circ$ for the two energy intervals adjacent to $\omega_m$ have comparable standard deviations, whereas the standard deviation for all other values is much smaller (of the order of symbol sizes in Fig. \ref{rotexp}). This is generally found to be the case, i.e., the (typically relatively few) values of $\vec S_\circ$ for which it is most important, are affected by the largest uncertainty. Thus, at least for the roton wave vector MaxEnt yields a reasonably accurate estimate of the position and the width of the peak, with some remaining uncertainty regarding its height.
\begin{figure}[t]
\centering
\includegraphics[width=0.47\textwidth]{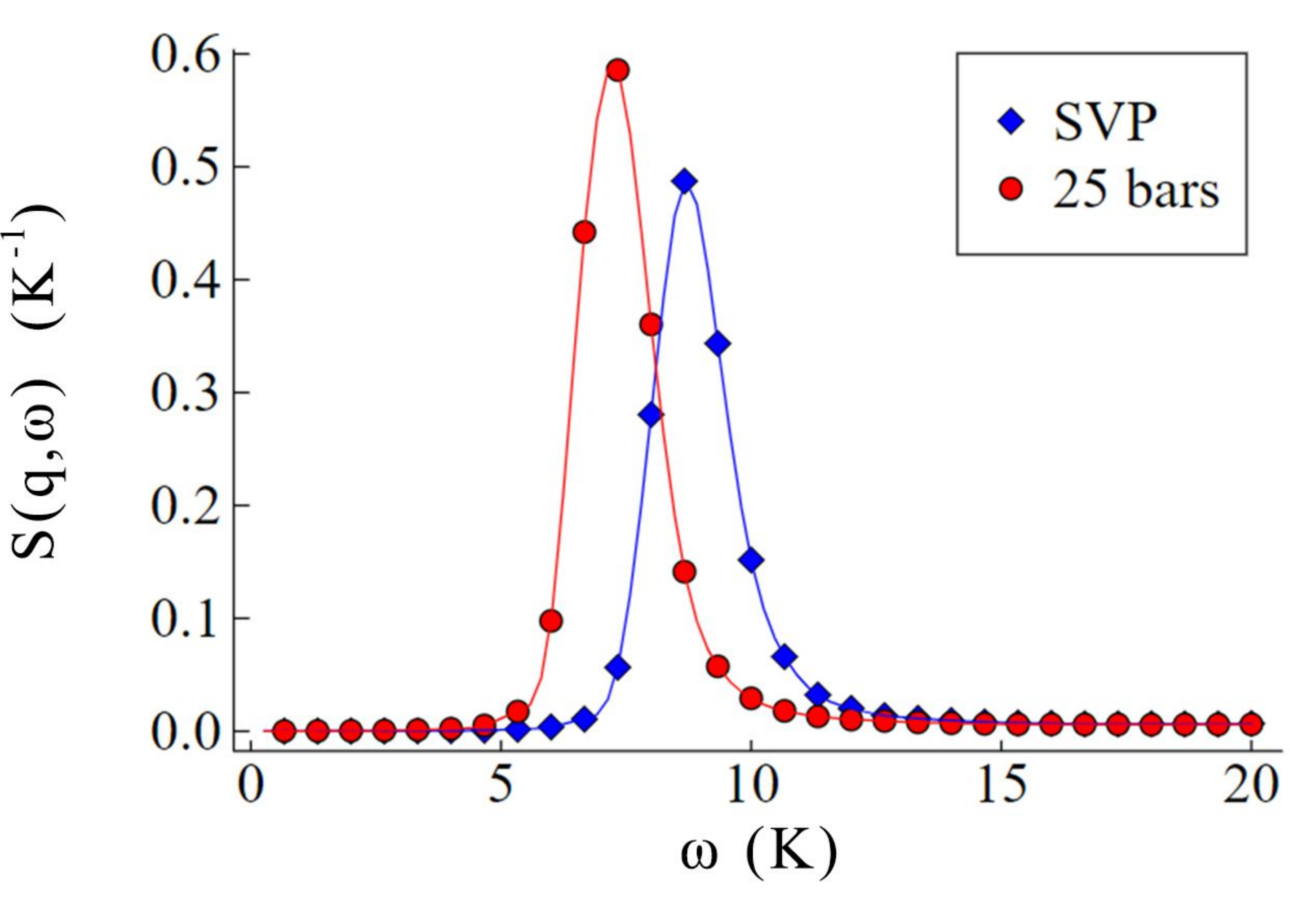}
\caption{{\em Color online}.  $S(\vec{q},\omega)$, defined as $\vec S_\circ$ (eq. \ref{optimal}) and computed as illustrated in the text, for superfluid $^4$He at $T=1$ K for the roton wave vector at SVP (diamonds, $q=1.963$ \AA$^{-1}$) and at 25 bars (circles, $q=2.081$ \AA$^{-1}$). Statistical errors on $S(\vec{q},\omega)$ are comparable to the sizes of the symbols for both curves.}
\label{rotpress}     
\end{figure}
\\ \indent 
It is interesting to note that, despite the uncertainty, nevertheless relative comparisons of data obtained with the procedure illustrated here are still  meaningful. For example, Fig. \ref{rotpress} shows $S({\bf q},\omega)$ for the roton wave vector at two different pressures, namely SVP and 25 bars. The roton minimum shifts from $\sim 1.9$ \AA$^{-1}$ at SVP to $\sim 2.1$ \AA$^{-1}$ at 25 bars \cite{pearce}. Our results show that the position of the peak moves to lower energy and the peak itself gains strength, in remarkable quantitative agreement with experimental observation \cite{gibbs}. 
\begin{figure}[t]
\centering
\includegraphics[width=0.47\textwidth]{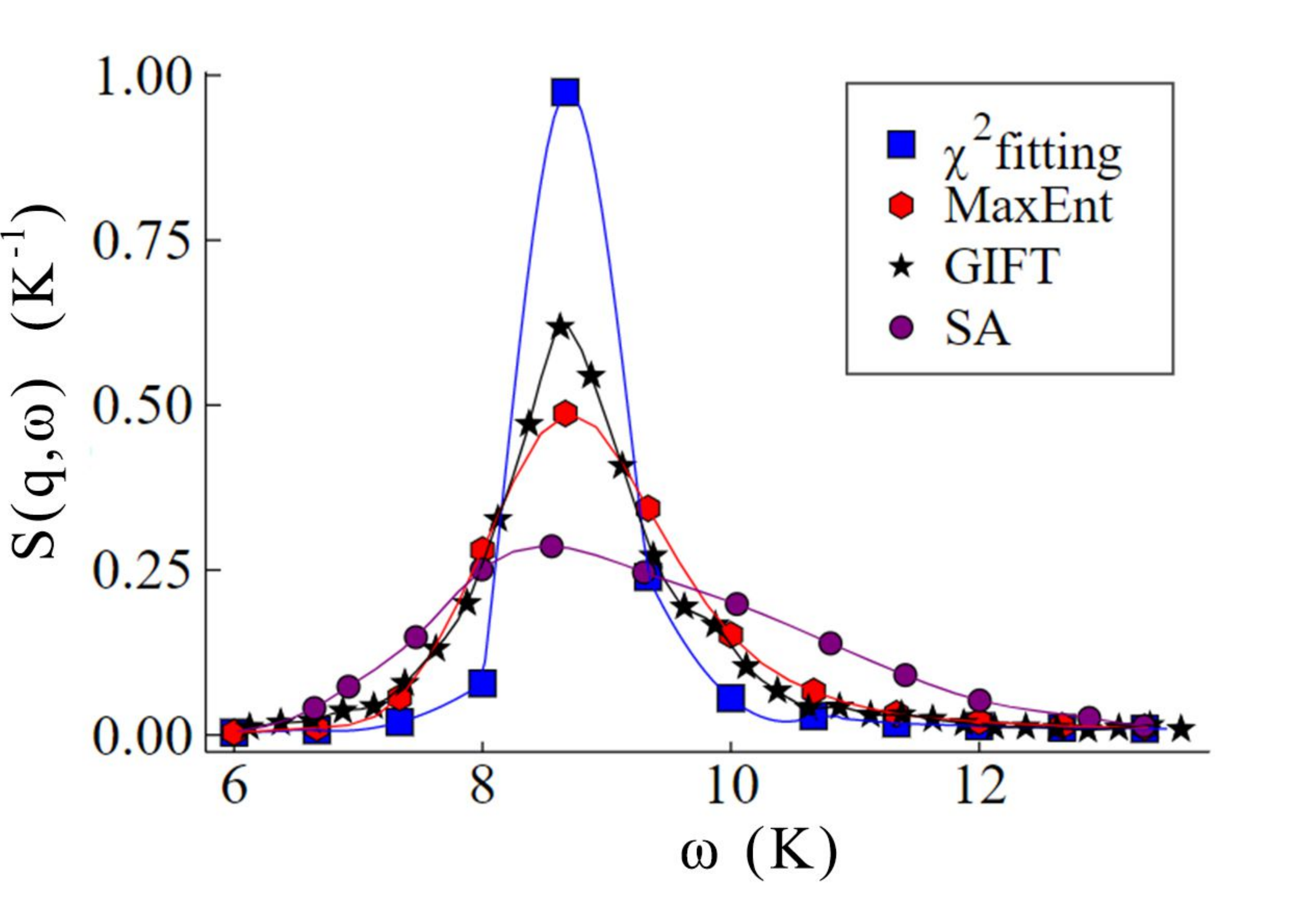}
\caption{{\em Color online}. $S(\vec{q},\omega)$ in superfluid $^4$He for the roton wave vector ($q=1.963$  \AA$^{-1}$) calculated through the inversion of QMC data based on four different methods. Hexagons show the result of the inversion using MaxEnt (eq. \ref{optimal}), whereas squares show that with $\alpha=0$ (which amounts to standard $\chi^2$ fitting).  Stars show the result of the inversion using GIFT \cite{gift} for the wave vector $q=1.977$  \AA$^{-1}$ at $T=0$ K.  Dark circles show the result of $\chi^2$-minimization using simulated annealing (SA) \cite{ferre} for the wave vector $q=1.91$  \AA$^{-1}$ at $T=0.8$ K.}\label{rotall}
\end{figure}
\\ \indent 
In Fig. \ref{rotall}, we compare our results with those of other authors who made use of  different approaches (not based on MaxEnt) to tackle the inversion of QMC data  \cite{private_gift}. 
The wave vectors are not identical but are reasonably close to the roton minimum in all cases; all calculations are carried out in the low temperature limit (see caption of Fig. \ref {rotall} for details).
There is nearly perfect agreement between our image and that of Ref. \onlinecite{gift}, especially if the standard deviation of our result is taken into account. On the other hand, the spectral image obtained in Ref. \onlinecite{ferre} is much broader, with a significantly lower peak. It is interesting to compare these curves with that arising from $\chi^2$-fitting carried out in the context of our procedure, namely by simply setting $\alpha=0$. In this case, the average value of $\chi^2$ is $\sim 0.2\ L$, i.e.,  slightly lower than that obtained with finite $\alpha$. However, as can be seen in Fig. \ref{rotall}, the peak is significantly higher (in fact its height exceeds that of the experimental result by almost a factor two) and also narrower than what is observed experimentally. This is consistent with the general notion that ``brute force'' $\chi^2$ minimization, while yielding sharp features, is all too likely to result in unphysical behavior. The use of the entropic prior emphasizes the contribution from smoother images (still consistent with the QMC data), which in this case results in better agreement with experiment.
\begin{figure}[h]
\centering
\includegraphics[width=0.47\textwidth]{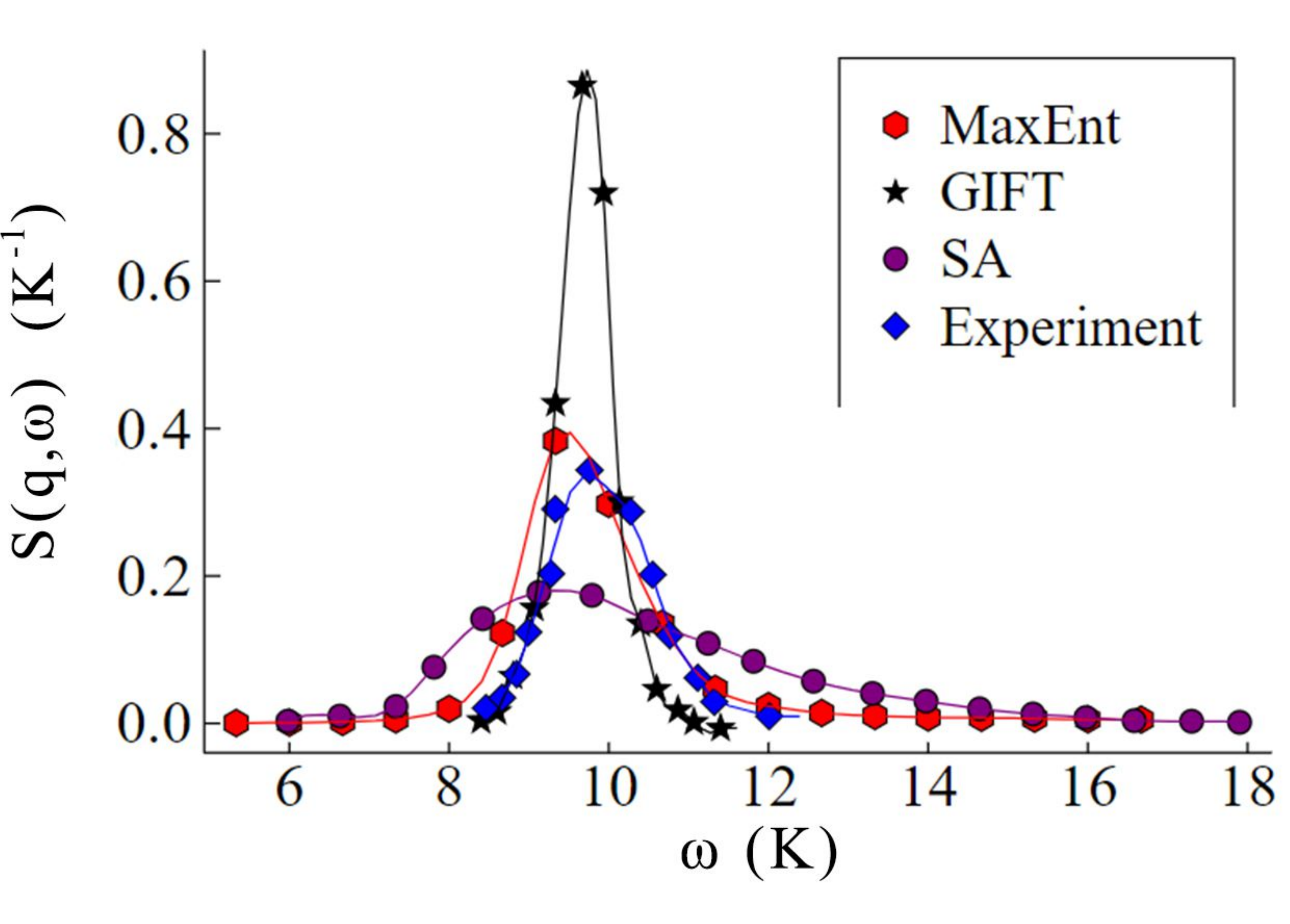}
\caption{{\em Color online}. $S(\vec{q},\omega)$ in superfluid $^4$He for the wave vector $q=1.756$  \AA$^{-1}$ calculated through the inversion of QMC data based on three different methods. Hexagons show the result of the inversion using MaxEnt (eq. \ref{optimal}). Stars show the result of the inversion using GIFT \cite{gift} for the wave vector $q=1.755$  \AA$^{-1}$ at $T=0$ K.  Dark circles show the result of $\chi^2$-minimization using simulated annealing (SA) \cite{ferre} for the wave vector $q=1.76$  \AA$^{-1}$ at $T=1.2$ K. Diamonds show experimental data from Ref. \onlinecite{andersen1994} (only the coherent part is shown) at $T=1.3$  K for the wave vector $q=1.70$  \AA$^{-1}$.}\label{q1_7}
\end{figure}
\\ \indent
Let us now consider a second wave vector, namely $q=1.756$  \AA$^{-1}$. In Fig. \ref{q1_7}, we compare again the result of our MaxEnt inversion with those of Refs. \onlinecite{gift,ferre}, as well as experimental data from Ref. \onlinecite{andersen1994}. Our procedure yields a spectral image in much closer agreement with experiment than the other two. In particular, both the shape of the curve and the location of the main peak are in excellent agreement with experiment, taking into account the slight difference in wave vectors \cite{donnelly} and the resolution of our spectral image. 
On the other hand, the spectral image reported in Ref. \onlinecite{ferre} is once again much too broad compared to the experimentally observed one, while that of Ref. \onlinecite{gift} is considerably sharper. 
 \begin{figure}[h]
\centering
\includegraphics[width=0.47\textwidth]{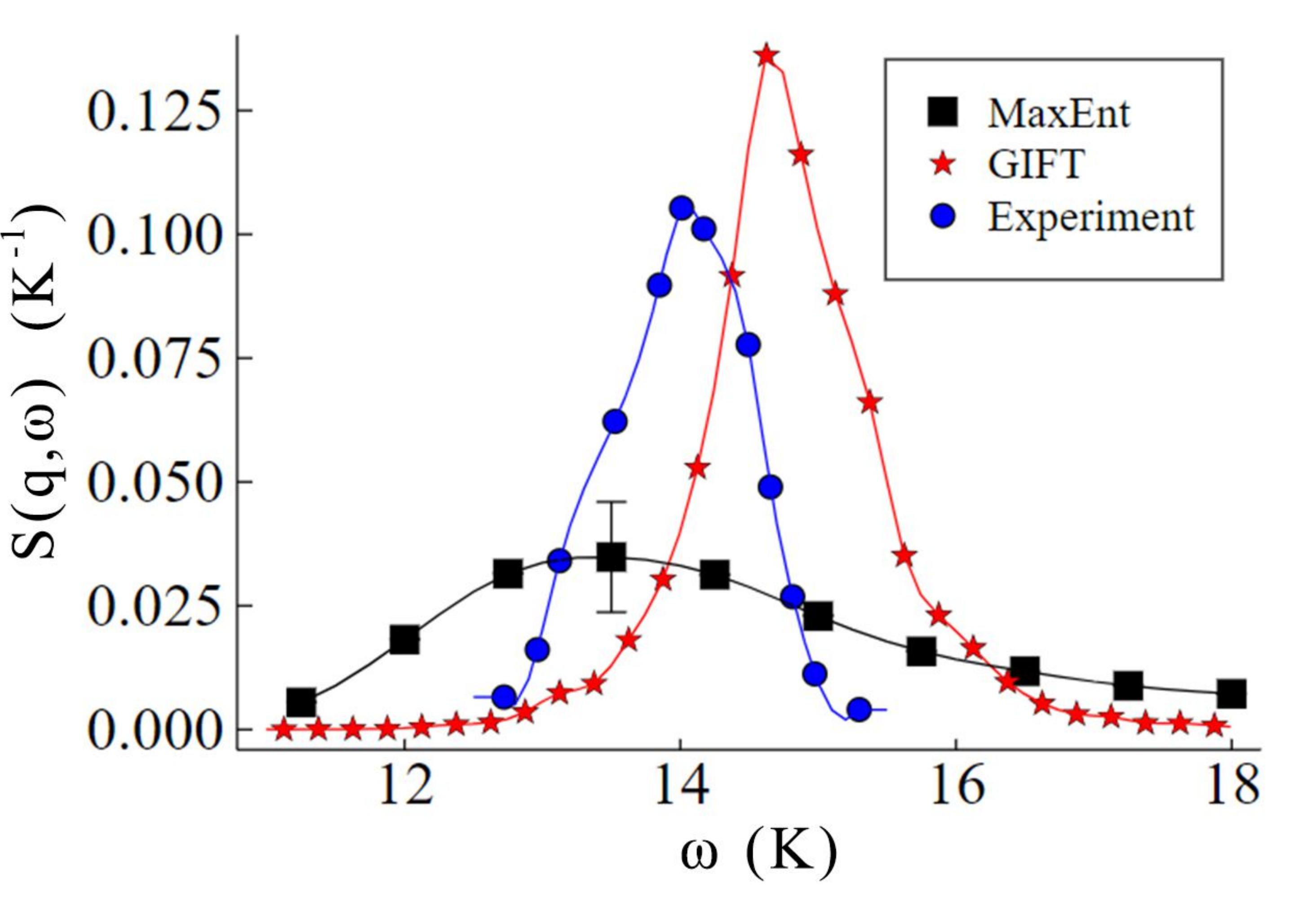}
\caption{{\em Color online}. $S(\vec{q},\omega)$ in superfluid $^4$He for the maxon wave vector ($q=1.075$  \AA$^{-1}$).  Squares show the result calculated through the inversion of QMC data based on MaxEnt (eq. \ref{optimal}). Stars show the result of the inversion using GIFT \cite{gift,private_gift}, calculated for the wave vector $q=1.107$  \AA$^{-1}$ at $T=0$ K. Circles show experimental data from Ref. \onlinecite{andersen1994} (only the coherent part is shown) at $T=1.3$  K for the wave vector $q=1.10$  \AA$^{-1}$. Statistical errors are of the order of the symbol sizes. The error bar on the square data point close to the peak represents a typical standard deviation.}
\label{maxall}     
\end{figure}
\\ \indent
Finally, let us examine results at a third wave vector, namely $q=1.075$ \AA$^{-1}$, which corresponds to the maxon.   In this case,  our spectral image features a single peak, which is however nowhere near as sharp as in the experimentally observed dynamic structure factor \cite{andersen1994}, as shown in Fig. \ref{maxall}. The considerably greater difficulty in extracting sharp features for this wave vector is a direct consequence of the behavior in imaginary time for $F({\bf q},\tau)$ , namely the much faster decay in the maxon case (Fig. \ref{f1}). Indeed, we find that the difficulty of reconstructing $S({\bf q},\omega)$ from QMC data is particularly severe for wave vectors near the maxon. For our procedure to recover sharp features at this wave vector, it appears that the underlying QMC data should possess errors that are significantly smaller than those which we could achieve within this project. This illustrates the difficulty of an a priori, even semi-quantitative assessment of the required precision of the QMC data.
\\ \indent 
Interestingly, the procedure illustrated in Ref. \onlinecite{gift} does yield a sharp peak in this case as well, of width comparable to that of the experimental image, and $\sim 30$\% greater height (data from Ref. \onlinecite{ferre} for this wave vector were not available). However, the position of the peak itself is off, compared to experiment, by roughly as much as that estimated by MaxEnt (in the case of GIFT the peak is detected at higher energy). Thus, although the shape of the GIFT image is certainly closer to the experimental result, in quantitative terms (e.g., position of the peak and area in the experimentally observed peak region), a comparison between the two results may not be so straightforward; in particular, one ought to think of situations in which this procedure is to be used in a {\em predictive} way, i.e., no experimental data are available for comparison.
\\ \indent 
Thus, we conclude that for this particular wave vector the precision required in the QMC data, in order to achieve a spectral image reconstruction of quality comparable to that of the other two wave vectors, is significantly greater than that afforded by the computational resources available to this project. It is incorrect to attribute the lack of sharpness of the reconstructed spectral image in this case to the inversion scheme utilized, which proves equally or more effective than the alternatives at other wave vectors.
\section{Conclusions}\label{conc}
We have revisited the use of MaxEnt to extract the dynamic structure factor of superfluid $^4$He from imaginary-time density correlation functions computed by QMC. This method was first applied to this problem over two decades ago, yielding results that were deemed ``only qualitatively interesting'', as the sharper features of the experimentally measured spectra were not fully recovered. In recent years, alternative schemes \cite{gift,ferre} have been proposed to tackle the same problem; although they are based on different numerical optimization strategies, these schemes ultimately amount to $\chi^2$-fitting.
\\ \indent 
We adopted in this work a procedure similar to that first proposed in Ref. \onlinecite{massimo1996}, i.e, we performed a random walk in the space of spectral images, using the entropic prior in the context of ``classic'' MaxEnt. Our study benefits from the availability  of new QMC data obtained using state-of-the-art techniques and obviously far more powerful computing resources than those available two decades ago. The accuracy of our QMC data is, to the best of our determination, comparable to that of the data used in Refs. \onlinecite{gift,ferre}. \\ \indent 
Our spectral images are of quality at least comparable (and often superior) to that of those yielded by the methods proposed in Refs. \onlinecite{gift,ferre}. In particular, spectral images provided in Ref. \onlinecite{ferre} are too broad, and compare poorly to experiment, whereas those of Ref. \onlinecite{gift} are at times much sharper than the experimental ones.
\\ \indent 
We show that the use of the entropic prior does {\em not} cause the reconstructed spectral images to be unphysically smooth and featureless. Rather, it is the precision of the underlying QMC data that determines by itself whether the reconstructed spectra should display sharp peaks or not.
In general, the elimination of the entropy from the inversion process indeed promotes sharper features, but we argue that that often comes at the expense of accuracy, as 
such sharpness is ultimately not warranted by the data. This means that some sharp features might appear at incorrect locations, or even be downright spurious. One is therefore left with no real justification to choose a ``sharper'' over a more conservative, smoother image, if both are consistent with the data ({\em a posteriori} validation based on agreement with available experiments for one particular physical system being a dubious criterion to compare different methodologies).  

\section*{acknowledgments}
This work was supported by the Natural Sciences and Engineering Research Council of Canada. One of us (MB) wishes to acknowledge the hospitality of the International Centre for Theoretical Physics in Trieste, Italy, where parts of the research work were carried out. The authors thank D. E. Galli for providing GIFT data. Useful conversations with S. Moroni are also gratefully acknowledged.

\bibliography{biblio}

\begin{thebibliography}{47}%
\makeatletter
\providecommand \@ifxundefined [1]{%
 \@ifx{#1\undefined}
}%
\providecommand \@ifnum [1]{%
 \ifnum #1\expandafter \@firstoftwo
 \else \expandafter \@secondoftwo
 \fi
}%
\providecommand \@ifx [1]{%
 \ifx #1\expandafter \@firstoftwo
 \else \expandafter \@secondoftwo
 \fi
}%
\providecommand \natexlab [1]{#1}%
\providecommand \enquote  [1]{``#1''}%
\providecommand \bibnamefont  [1]{#1}%
\providecommand \bibfnamefont [1]{#1}%
\providecommand \citenamefont [1]{#1}%
\providecommand \href@noop [0]{\@secondoftwo}%
\providecommand \href [0]{\begingroup \@sanitize@url \@href}%
\providecommand \@href[1]{\@@startlink{#1}\@@href}%
\providecommand \@@href[1]{\endgroup#1\@@endlink}%
\providecommand \@sanitize@url [0]{\catcode `\\12\catcode `\$12\catcode
  `\&12\catcode `\#12\catcode `\^12\catcode `\_12\catcode `\%12\relax}%
\providecommand \@@startlink[1]{}%
\providecommand \@@endlink[0]{}%
\providecommand \url  [0]{\begingroup\@sanitize@url \@url }%
\providecommand \@url [1]{\endgroup\@href {#1}{\urlprefix }}%
\providecommand \urlprefix  [0]{URL }%
\providecommand \Eprint [0]{\href }%
\providecommand \doibase [0]{http://dx.doi.org/}%
\providecommand \selectlanguage [0]{\@gobble}%
\providecommand \bibinfo  [0]{\@secondoftwo}%
\providecommand \bibfield  [0]{\@secondoftwo}%
\providecommand \translation [1]{[#1]}%
\providecommand \BibitemOpen [0]{}%
\providecommand \bibitemStop [0]{}%
\providecommand \bibitemNoStop [0]{.\EOS\space}%
\providecommand \EOS [0]{\spacefactor3000\relax}%
\providecommand \BibitemShut  [1]{\csname bibitem#1\endcsname}%
\let\auto@bib@innerbib\@empty
\bibitem [{\citenamefont {Boninsegni}\ \emph {et~al.}(2006)\citenamefont
  {Boninsegni}, \citenamefont {Prokof'ev},\ and\ \citenamefont
  {Svistunov}}]{worm}%
  \BibitemOpen
  \bibfield  {author} {\bibinfo {author} {\bibfnamefont {M.}~\bibnamefont
  {Boninsegni}}, \bibinfo {author} {\bibfnamefont {N.~V.}\ \bibnamefont
  {Prokof'ev}}, \ and\ \bibinfo {author} {\bibfnamefont {B.~V.}\ \bibnamefont
  {Svistunov}},\ }\href@noop {} {\bibfield  {journal} {\bibinfo  {journal}
  {Phys. Rev. E}\ }\textbf {\bibinfo {volume} {74}},\ \bibinfo {pages} {036701}
  (\bibinfo {year} {2006})}\BibitemShut {NoStop}%
\bibitem [{\citenamefont {Saccani}\ \emph {et~al.}(2012)\citenamefont
  {Saccani}, \citenamefont {Moroni},\ and\ \citenamefont
  {Boninsegni}}]{saccani}%
  \BibitemOpen
  \bibfield  {author} {\bibinfo {author} {\bibfnamefont {S.}~\bibnamefont
  {Saccani}}, \bibinfo {author} {\bibfnamefont {S.}~\bibnamefont {Moroni}}, \
  and\ \bibinfo {author} {\bibfnamefont {M.}~\bibnamefont {Boninsegni}},\
  }\href@noop {} {\bibfield  {journal} {\bibinfo  {journal} {Phys. Rev. Lett.}\
  }\textbf {\bibinfo {volume} {108}},\ \bibinfo {pages} {175301} (\bibinfo
  {year} {2012})}\BibitemShut {NoStop}%
\bibitem [{jus()}]{justsoyouknow}%
  \BibitemOpen
  \href@noop {} {}\bibinfo {note} {This should be regarded {\em not} as a
  limitation, but rather as a {\em quality} of the RS, as one ought not ascribe
  any physical significance to sharp, distinct features that could be spurious,
  in conformity with the accepted ``Occam's razor'' principle; credence should
  be lent only to those sharp features that remain consistently robust as the
  quality of the underlying data improves.}\BibitemShut {Stop}%
\bibitem [{\citenamefont {Papoulis}(1990)}]{probstat}%
  \BibitemOpen
  \bibfield  {author} {\bibinfo {author} {\bibfnamefont {A.}~\bibnamefont
  {Papoulis}},\ }\href@noop {} {\emph {\bibinfo {title} {Probability and
  Statistics}}}\ (\bibinfo  {publisher} {Prentice Hall},\ \bibinfo {address}
  {New York},\ \bibinfo {year} {1990})\BibitemShut {NoStop}%
\bibitem [{\citenamefont {Jarrel}\ and\ \citenamefont
  {Gubernatis}(1996)}]{bayesian}%
  \BibitemOpen
  \bibfield  {author} {\bibinfo {author} {\bibfnamefont {M.}~\bibnamefont
  {Jarrel}}\ and\ \bibinfo {author} {\bibfnamefont {J.}~\bibnamefont
  {Gubernatis}},\ }\href@noop {} {\bibfield  {journal} {\bibinfo  {journal}
  {Phys. Rep.}\ }\textbf {\bibinfo {volume} {269}},\ \bibinfo {pages} {133}
  (\bibinfo {year} {1996})}\BibitemShut {NoStop}%
\bibitem [{\citenamefont {Silver}\ \emph {et~al.}(1990)\citenamefont {Silver},
  \citenamefont {Gubernatis}, \citenamefont {Sivia},\ and\ \citenamefont
  {Jarrell}}]{gubernatis}%
  \BibitemOpen
  \bibfield  {author} {\bibinfo {author} {\bibfnamefont {R.~N.}\ \bibnamefont
  {Silver}}, \bibinfo {author} {\bibfnamefont {J.~E.}\ \bibnamefont
  {Gubernatis}}, \bibinfo {author} {\bibfnamefont {D.~S.}\ \bibnamefont
  {Sivia}}, \ and\ \bibinfo {author} {\bibfnamefont {M.}~\bibnamefont
  {Jarrell}},\ }\href@noop {} {\bibfield  {journal} {\bibinfo  {journal} {Phys.
  Rev. Lett.}\ }\textbf {\bibinfo {volume} {65}},\ \bibinfo {pages} {496}
  (\bibinfo {year} {1990})}\BibitemShut {NoStop}%
\bibitem [{\citenamefont {White}(1991)}]{white}%
  \BibitemOpen
  \bibfield  {author} {\bibinfo {author} {\bibfnamefont {S.~R.}\ \bibnamefont
  {White}},\ }\href@noop {} {\bibfield  {journal} {\bibinfo  {journal} {Phys.
  Rev. B}\ }\textbf {\bibinfo {volume} {44}},\ \bibinfo {pages} {4670}
  (\bibinfo {year} {1991})}\BibitemShut {NoStop}%
\bibitem [{\citenamefont {Makivi\'c}\ and\ \citenamefont
  {Jarrell}(1992)}]{makivic}%
  \BibitemOpen
  \bibfield  {author} {\bibinfo {author} {\bibfnamefont {M.}~\bibnamefont
  {Makivi\'c}}\ and\ \bibinfo {author} {\bibfnamefont {M.}~\bibnamefont
  {Jarrell}},\ }\href@noop {} {\bibfield  {journal} {\bibinfo  {journal} {Phys.
  Rev. Lett.}\ }\textbf {\bibinfo {volume} {68}},\ \bibinfo {pages} {1770}
  (\bibinfo {year} {1992})}\BibitemShut {NoStop}%
\bibitem [{\citenamefont {Bulut}\ \emph {et~al.}(1994)\citenamefont {Bulut},
  \citenamefont {Scalapino},\ and\ \citenamefont {White}}]{bulut}%
  \BibitemOpen
  \bibfield  {author} {\bibinfo {author} {\bibfnamefont {N.}~\bibnamefont
  {Bulut}}, \bibinfo {author} {\bibfnamefont {D.~J.}\ \bibnamefont
  {Scalapino}}, \ and\ \bibinfo {author} {\bibfnamefont {S.~R.}\ \bibnamefont
  {White}},\ }\href@noop {} {\bibfield  {journal} {\bibinfo  {journal} {Phys.
  Rev. Lett.}\ }\textbf {\bibinfo {volume} {72}},\ \bibinfo {pages} {705}
  (\bibinfo {year} {1994})}\BibitemShut {NoStop}%
\bibitem [{\citenamefont {Preuss}\ \emph {et~al.}(1994)\citenamefont {Preuss},
  \citenamefont {Muramatsu}, \citenamefont {von~der Linden}, \citenamefont
  {Dieterich}, \citenamefont {Assaad},\ and\ \citenamefont {Hanke}}]{preuss}%
  \BibitemOpen
  \bibfield  {author} {\bibinfo {author} {\bibfnamefont {R.}~\bibnamefont
  {Preuss}}, \bibinfo {author} {\bibfnamefont {A.}~\bibnamefont {Muramatsu}},
  \bibinfo {author} {\bibfnamefont {W.}~\bibnamefont {von~der Linden}},
  \bibinfo {author} {\bibfnamefont {P.}~\bibnamefont {Dieterich}}, \bibinfo
  {author} {\bibfnamefont {F.~F.}\ \bibnamefont {Assaad}}, \ and\ \bibinfo
  {author} {\bibfnamefont {W.}~\bibnamefont {Hanke}},\ }\href@noop {}
  {\bibfield  {journal} {\bibinfo  {journal} {Phys. Rev. Lett.}\ }\textbf
  {\bibinfo {volume} {73}},\ \bibinfo {pages} {732} (\bibinfo {year}
  {1994})}\BibitemShut {NoStop}%
\bibitem [{\citenamefont {Boninsegni}\ and\ \citenamefont
  {Ceperley}(1996)}]{massimo1996}%
  \BibitemOpen
  \bibfield  {author} {\bibinfo {author} {\bibfnamefont {M.}~\bibnamefont
  {Boninsegni}}\ and\ \bibinfo {author} {\bibfnamefont {D.~M.}\ \bibnamefont
  {Ceperley}},\ }\href@noop {} {\bibfield  {journal} {\bibinfo  {journal} {J.
  Low Temp. Phys.}\ }\textbf {\bibinfo {volume} {104}},\ \bibinfo {pages} {339}
  (\bibinfo {year} {1996})}\BibitemShut {NoStop}%
\bibitem [{\citenamefont {Sandvik}(1998)}]{sandvik}%
  \BibitemOpen
  \bibfield  {author} {\bibinfo {author} {\bibfnamefont {A.~W.}\ \bibnamefont
  {Sandvik}},\ }\href@noop {} {\bibfield  {journal} {\bibinfo  {journal} {Phys.
  Rev. B}\ }\textbf {\bibinfo {volume} {57}},\ \bibinfo {pages} {10287}
  (\bibinfo {year} {1998})}\BibitemShut {NoStop}%
\bibitem [{\citenamefont {Sylju{\aa}sen}(2008)}]{sylvjuasen}%
  \BibitemOpen
  \bibfield  {author} {\bibinfo {author} {\bibfnamefont {O.~F.}\ \bibnamefont
  {Sylju{\aa}sen}},\ }\href@noop {} {\bibfield  {journal} {\bibinfo  {journal}
  {Phys. Rev. B}\ }\textbf {\bibinfo {volume} {78}},\ \bibinfo {pages} {174429}
  (\bibinfo {year} {2008})}\BibitemShut {NoStop}%
\bibitem [{\citenamefont {Mishchenko}\ \emph {et~al.}(2000)\citenamefont
  {Mishchenko}, \citenamefont {Prokof'ev}, \citenamefont {Sakamoto},\ and\
  \citenamefont {Svistunov}}]{mishenko}%
  \BibitemOpen
  \bibfield  {author} {\bibinfo {author} {\bibfnamefont {A.~S.}\ \bibnamefont
  {Mishchenko}}, \bibinfo {author} {\bibfnamefont {N.~V.}\ \bibnamefont
  {Prokof'ev}}, \bibinfo {author} {\bibfnamefont {A.}~\bibnamefont {Sakamoto}},
  \ and\ \bibinfo {author} {\bibfnamefont {B.~V.}\ \bibnamefont {Svistunov}},\
  }\href@noop {} {\bibfield  {journal} {\bibinfo  {journal} {Phys. Rev. B}\
  }\textbf {\bibinfo {volume} {62}},\ \bibinfo {pages} {6317} (\bibinfo {year}
  {2000})}\BibitemShut {NoStop}%
\bibitem [{\citenamefont {Reichman}\ and\ \citenamefont
  {Rabani}(2009)}]{reichman}%
  \BibitemOpen
  \bibfield  {author} {\bibinfo {author} {\bibfnamefont {D.~R.}\ \bibnamefont
  {Reichman}}\ and\ \bibinfo {author} {\bibfnamefont {E.}~\bibnamefont
  {Rabani}},\ }\href@noop {} {\bibfield  {journal} {\bibinfo  {journal} {J.
  Chem. Phys.}\ }\textbf {\bibinfo {volume} {131}},\ \bibinfo {pages} {054502}
  (\bibinfo {year} {2009})}\BibitemShut {NoStop}%
\bibitem [{\citenamefont {Fuchs}\ \emph {et~al.}(2010)\citenamefont {Fuchs},
  \citenamefont {Pruschke},\ and\ \citenamefont {Jarrell}}]{fuchs}%
  \BibitemOpen
  \bibfield  {author} {\bibinfo {author} {\bibfnamefont {S.}~\bibnamefont
  {Fuchs}}, \bibinfo {author} {\bibfnamefont {T.}~\bibnamefont {Pruschke}}, \
  and\ \bibinfo {author} {\bibfnamefont {M.}~\bibnamefont {Jarrell}},\
  }\href@noop {} {\bibfield  {journal} {\bibinfo  {journal} {Phys. Rev. E}\
  }\textbf {\bibinfo {volume} {81}},\ \bibinfo {pages} {056701} (\bibinfo
  {year} {2010})}\BibitemShut {NoStop}%
\bibitem [{\citenamefont {Vitali}\ \emph {et~al.}(2010)\citenamefont {Vitali},
  \citenamefont {Rossi}, \citenamefont {Reatto},\ and\ \citenamefont
  {Galli}}]{gift}%
  \BibitemOpen
  \bibfield  {author} {\bibinfo {author} {\bibfnamefont {E.}~\bibnamefont
  {Vitali}}, \bibinfo {author} {\bibfnamefont {M.}~\bibnamefont {Rossi}},
  \bibinfo {author} {\bibfnamefont {L.}~\bibnamefont {Reatto}}, \ and\ \bibinfo
  {author} {\bibfnamefont {D.~E.}\ \bibnamefont {Galli}},\ }\href@noop {}
  {\bibfield  {journal} {\bibinfo  {journal} {Phys. Rev. B}\ }\textbf {\bibinfo
  {volume} {82}},\ \bibinfo {pages} {174510} (\bibinfo {year}
  {2010})}\BibitemShut {NoStop}%
\bibitem [{\citenamefont {Ferr\'e}\ and\ \citenamefont
  {Boronat}(2016)}]{ferre}%
  \BibitemOpen
  \bibfield  {author} {\bibinfo {author} {\bibfnamefont {G.}~\bibnamefont
  {Ferr\'e}}\ and\ \bibinfo {author} {\bibfnamefont {J.}~\bibnamefont
  {Boronat}},\ }\href@noop {} {\bibfield  {journal} {\bibinfo  {journal} {Phys.
  Rev. B}\ }\textbf {\bibinfo {volume} {93}},\ \bibinfo {pages} {104510}
  (\bibinfo {year} {2016})}\BibitemShut {NoStop}%
\bibitem [{dif()}]{diff}%
  \BibitemOpen
  \href@noop {} {}\bibinfo {note} {The main difference between the approaches
  proposed in Refs. \onlinecite {gift} and \onlinecite{ferre} is the numerical
  methodology adopted to identify the optimal image, i.e., to minimize the
  value of $\chi^2$.}\BibitemShut {Stop}%
\bibitem [{\citenamefont {Caffarel}\ and\ \citenamefont
  {Ceperley}(1992)}]{caffarel}%
  \BibitemOpen
  \bibfield  {author} {\bibinfo {author} {\bibfnamefont {M.}~\bibnamefont
  {Caffarel}}\ and\ \bibinfo {author} {\bibfnamefont {D.~M.}\ \bibnamefont
  {Ceperley}},\ }\href@noop {} {\bibfield  {journal} {\bibinfo  {journal} {J.
  Chem. Phys.}\ }\textbf {\bibinfo {volume} {97}},\ \bibinfo {pages} {8415}
  (\bibinfo {year} {1992})}\BibitemShut {NoStop}%
\bibitem [{\citenamefont {Boninsegni}\ and\ \citenamefont
  {Manousakis}(1992)}]{bm}%
  \BibitemOpen
  \bibfield  {author} {\bibinfo {author} {\bibfnamefont {M.}~\bibnamefont
  {Boninsegni}}\ and\ \bibinfo {author} {\bibfnamefont {E.}~\bibnamefont
  {Manousakis}},\ }\href@noop {} {\bibfield  {journal} {\bibinfo  {journal}
  {Phys. Rev. B}\ }\textbf {\bibinfo {volume} {46}},\ \bibinfo {pages} {560}
  (\bibinfo {year} {1992})}\BibitemShut {NoStop}%
\bibitem [{\citenamefont {Aziz}\ \emph {et~al.}(1979)\citenamefont {Aziz},
  \citenamefont {Nain}, \citenamefont {Carley}, \citenamefont {Taylor},\ and\
  \citenamefont {McConville}}]{aziz79}%
  \BibitemOpen
  \bibfield  {author} {\bibinfo {author} {\bibfnamefont {R.~A.}\ \bibnamefont
  {Aziz}}, \bibinfo {author} {\bibfnamefont {V.~P.~S.}\ \bibnamefont {Nain}},
  \bibinfo {author} {\bibfnamefont {J.~S.}\ \bibnamefont {Carley}}, \bibinfo
  {author} {\bibfnamefont {W.~L.}\ \bibnamefont {Taylor}}, \ and\ \bibinfo
  {author} {\bibfnamefont {G.~T.}\ \bibnamefont {McConville}},\ }\href@noop {}
  {\bibfield  {journal} {\bibinfo  {journal} {J. Chem. Phys.}\ }\textbf
  {\bibinfo {volume} {70}},\ \bibinfo {pages} {4330} (\bibinfo {year}
  {1979})}\BibitemShut {NoStop}%
\bibitem [{\citenamefont {Moroni}\ \emph {et~al.}(2000)\citenamefont {Moroni},
  \citenamefont {Pederiva}, \citenamefont {Fantoni},\ and\ \citenamefont
  {Boninsegni}}]{mpfb}%
  \BibitemOpen
  \bibfield  {author} {\bibinfo {author} {\bibfnamefont {S.}~\bibnamefont
  {Moroni}}, \bibinfo {author} {\bibfnamefont {F.}~\bibnamefont {Pederiva}},
  \bibinfo {author} {\bibfnamefont {S.}~\bibnamefont {Fantoni}}, \ and\
  \bibinfo {author} {\bibfnamefont {M.}~\bibnamefont {Boninsegni}},\
  }\href@noop {} {\bibfield  {journal} {\bibinfo  {journal} {Phys. Rev. Lett.}\
  }\textbf {\bibinfo {volume} {84}},\ \bibinfo {pages} {2650} (\bibinfo {year}
  {2000})}\BibitemShut {NoStop}%
\bibitem [{\citenamefont {Mezzacapo}\ and\ \citenamefont
  {Boninsegni}(2006)}]{mezz1}%
  \BibitemOpen
  \bibfield  {author} {\bibinfo {author} {\bibfnamefont {F.}~\bibnamefont
  {Mezzacapo}}\ and\ \bibinfo {author} {\bibfnamefont {M.}~\bibnamefont
  {Boninsegni}},\ }\href@noop {} {\bibfield  {journal} {\bibinfo  {journal}
  {Phys. Rev. Lett.}\ }\textbf {\bibinfo {volume} {97}},\ \bibinfo {pages}
  {045301} (\bibinfo {year} {2006})}\BibitemShut {NoStop}%
\bibitem [{\citenamefont {Mezzacapo}\ and\ \citenamefont
  {Boninsegni}(2007)}]{mezz2}%
  \BibitemOpen
  \bibfield  {author} {\bibinfo {author} {\bibfnamefont {F.}~\bibnamefont
  {Mezzacapo}}\ and\ \bibinfo {author} {\bibfnamefont {M.}~\bibnamefont
  {Boninsegni}},\ }\href@noop {} {\bibfield  {journal} {\bibinfo  {journal}
  {Phys. Rev. A}\ }\textbf {\bibinfo {volume} {75}},\ \bibinfo {pages} {033201}
  (\bibinfo {year} {2007})}\BibitemShut {NoStop}%
\bibitem [{\citenamefont {Glyde}(2018)}]{glyde}%
  \BibitemOpen
  \bibfield  {author} {\bibinfo {author} {\bibfnamefont {H.~R.}\ \bibnamefont
  {Glyde}},\ }\href@noop {} {\bibfield  {journal} {\bibinfo  {journal} {Rep.
  Prog. Phys.}\ }\textbf {\bibinfo {volume} {81}},\ \bibinfo {pages} {014501}
  (\bibinfo {year} {2018})}\BibitemShut {NoStop}%
\bibitem [{\citenamefont {Landau}(1941)}]{landau}%
  \BibitemOpen
  \bibfield  {author} {\bibinfo {author} {\bibfnamefont {L.~D.}\ \bibnamefont
  {Landau}},\ }\href@noop {} {\bibfield  {journal} {\bibinfo  {journal} {J.
  Phys. USSR}\ }\textbf {\bibinfo {volume} {5}},\ \bibinfo {pages} {71}
  (\bibinfo {year} {1941})}\BibitemShut {NoStop}%
\bibitem [{\citenamefont {Feynman}(1954)}]{feynman}%
  \BibitemOpen
  \bibfield  {author} {\bibinfo {author} {\bibfnamefont {R.~P.}\ \bibnamefont
  {Feynman}},\ }\href@noop {} {\bibfield  {journal} {\bibinfo  {journal} {Phys.
  Rev.}\ }\textbf {\bibinfo {volume} {94}},\ \bibinfo {pages} {262} (\bibinfo
  {year} {1954})}\BibitemShut {NoStop}%
\bibitem [{\citenamefont {Feynman}\ and\ \citenamefont {Cohen}(1956)}]{cohen}%
  \BibitemOpen
  \bibfield  {author} {\bibinfo {author} {\bibfnamefont {R.~P.}\ \bibnamefont
  {Feynman}}\ and\ \bibinfo {author} {\bibfnamefont {M.}~\bibnamefont
  {Cohen}},\ }\href@noop {} {\bibfield  {journal} {\bibinfo  {journal} {Phys.
  Rev.}\ }\textbf {\bibinfo {volume} {102}},\ \bibinfo {pages} {1189} (\bibinfo
  {year} {1956})}\BibitemShut {NoStop}%
\bibitem [{\citenamefont {Lovesey}(1986)}]{lovesey}%
  \BibitemOpen
  \bibfield  {author} {\bibinfo {author} {\bibfnamefont {S.~W.}\ \bibnamefont
  {Lovesey}},\ }\href@noop {} {\emph {\bibinfo {title} {Condensed matter
  physics : dynamic correlations}}}\ (\bibinfo  {publisher}
  {Benjamin/Cummings},\ \bibinfo {address} {Menlo Park, California},\ \bibinfo
  {year} {1986})\BibitemShut {NoStop}%
\bibitem [{\citenamefont {Boninsegni}(2005)}]{jltp}%
  \BibitemOpen
  \bibfield  {author} {\bibinfo {author} {\bibfnamefont {M.}~\bibnamefont
  {Boninsegni}},\ }\href@noop {} {\bibfield  {journal} {\bibinfo  {journal} {J.
  Low Temp. Phys.}\ }\textbf {\bibinfo {volume} {27}},\ \bibinfo {pages} {141}
  (\bibinfo {year} {2005})}\BibitemShut {NoStop}%
\bibitem [{\citenamefont {Donnelly}\ and\ \citenamefont
  {Barenghi}(1998)}]{barenghi}%
  \BibitemOpen
  \bibfield  {author} {\bibinfo {author} {\bibfnamefont {R.~J.}\ \bibnamefont
  {Donnelly}}\ and\ \bibinfo {author} {\bibfnamefont {C.~F.}\ \bibnamefont
  {Barenghi}},\ }\href@noop {} {\bibfield  {journal} {\bibinfo  {journal} {J.
  Phys. Chem. Ref. Data}\ }\textbf {\bibinfo {volume} {27}},\ \bibinfo {pages}
  {1217} (\bibinfo {year} {1998})}\BibitemShut {NoStop}%
\bibitem [{\citenamefont {Gibbs}\ \emph {et~al.}(1999)\citenamefont {Gibbs},
  \citenamefont {Andersen}, \citenamefont {Stirling},\ and\ \citenamefont
  {Schober}}]{gibbs}%
  \BibitemOpen
  \bibfield  {author} {\bibinfo {author} {\bibfnamefont {M.~R.}\ \bibnamefont
  {Gibbs}}, \bibinfo {author} {\bibfnamefont {K.~H.}\ \bibnamefont {Andersen}},
  \bibinfo {author} {\bibfnamefont {W.~G.}\ \bibnamefont {Stirling}}, \ and\
  \bibinfo {author} {\bibfnamefont {H.}~\bibnamefont {Schober}},\ }\href@noop
  {} {\bibfield  {journal} {\bibinfo  {journal} {J. Phys.: Condens. Matter}\
  }\textbf {\bibinfo {volume} {11}},\ \bibinfo {pages} {603} (\bibinfo {year}
  {1999})}\BibitemShut {NoStop}%
\bibitem [{\citenamefont {Dietrich}\ \emph {et~al.}(1972)\citenamefont
  {Dietrich}, \citenamefont {Graf}, \citenamefont {Huang},\ and\ \citenamefont
  {Passell}}]{dietrich1972}%
  \BibitemOpen
  \bibfield  {author} {\bibinfo {author} {\bibfnamefont {O.}~\bibnamefont
  {Dietrich}}, \bibinfo {author} {\bibfnamefont {E.}~\bibnamefont {Graf}},
  \bibinfo {author} {\bibfnamefont {C.}~\bibnamefont {Huang}}, \ and\ \bibinfo
  {author} {\bibfnamefont {L.}~\bibnamefont {Passell}},\ }\href@noop {}
  {\bibfield  {journal} {\bibinfo  {journal} {Phys. Rev. A}\ }\textbf {\bibinfo
  {volume} {5}},\ \bibinfo {pages} {1377} (\bibinfo {year} {1972})}\BibitemShut
  {NoStop}%
\bibitem [{\citenamefont {Ceperley}(1995)}]{rmp}%
  \BibitemOpen
  \bibfield  {author} {\bibinfo {author} {\bibfnamefont {D.~M.}\ \bibnamefont
  {Ceperley}},\ }\href@noop {} {\bibfield  {journal} {\bibinfo  {journal} {Rev.
  Mod. Phys.}\ }\textbf {\bibinfo {volume} {67}},\ \bibinfo {pages} {279}
  (\bibinfo {year} {1995})}\BibitemShut {NoStop}%
\bibitem [{\citenamefont {Flyvbjerg}\ and\ \citenamefont
  {Petersen}(1989)}]{flyvbjerg1989}%
  \BibitemOpen
  \bibfield  {author} {\bibinfo {author} {\bibfnamefont {H.}~\bibnamefont
  {Flyvbjerg}}\ and\ \bibinfo {author} {\bibfnamefont {H.~G.}\ \bibnamefont
  {Petersen}},\ }\href@noop {} {\bibfield  {journal} {\bibinfo  {journal} {J.
  Chem. Phys.}\ }\textbf {\bibinfo {volume} {91}},\ \bibinfo {pages} {461}
  (\bibinfo {year} {1989})}\BibitemShut {NoStop}%
\bibitem [{ts()}]{ts}%
  \BibitemOpen
  \href@noop {} {}\bibinfo {note} {As mentioned in the text, the value of the
  time step utilized in the QMC calculation is $\epsilon=1/640$ K$^{-1}$. There
  are therefore 320 ``time slices'' in the imaginary-time interval
  $0\le\tau\le\beta/2$, but because the fourth-order formula is adopted, only
  half of them are usable for computation of expectation values of observables.
  For details, see, for instance, Ref. \onlinecite{jiang}}\BibitemShut
  {NoStop}%
\bibitem [{not()}]{noteco}%
  \BibitemOpen
  \href@noop {} {}\bibinfo {note} {In principle, the diagonal approximation for
  $C$ is not justified, because QMC data at different imaginary-times are not
  generated independently, i.e., they are correlated. However, the diagonal
  approximation often allows for a more stable inversion, and in practice the
  use of the full covariance matrix does not yield any significant difference
  in the results. See, for instance, Ref.
  \onlinecite{massimo1996}.}\BibitemShut {Stop}%
\bibitem [{\citenamefont {Jaynes}(1957)}]{jaynes}%
  \BibitemOpen
  \bibfield  {author} {\bibinfo {author} {\bibfnamefont {E.~T.}\ \bibnamefont
  {Jaynes}},\ }\href@noop {} {\bibfield  {journal} {\bibinfo  {journal} {Phys.
  Rev.}\ }\textbf {\bibinfo {volume} {106}},\ \bibinfo {pages} {620} (\bibinfo
  {year} {1957})}\BibitemShut {NoStop}%
\bibitem [{zip()}]{zipo}%
  \BibitemOpen
  \href@noop {} {}\bibinfo {note} {Implicit in the definition (\ref{jay}) is
  the use of a ``flat'' default model, i.e., one making no {\em a priori}
  assumption on the shape of $S$.}\BibitemShut {Stop}%
\bibitem [{non()}]{noneg}%
  \BibitemOpen
  \href@noop {} {}\bibinfo {note} {Obviously, moves attempting to make any of
  $S_j$ or $\alpha$ negative are automatically rejected.}\BibitemShut {Stop}%
\bibitem [{\citenamefont {Andersen}\ \emph {et~al.}(1994)\citenamefont
  {Andersen}, \citenamefont {Stirling}, \citenamefont {Scherm}, \citenamefont
  {Stunault}, \citenamefont {Fak}, \citenamefont {Godfrin},\ and\ \citenamefont
  {Dianoux}}]{andersen1994}%
  \BibitemOpen
  \bibfield  {author} {\bibinfo {author} {\bibfnamefont {K.~H.}\ \bibnamefont
  {Andersen}}, \bibinfo {author} {\bibfnamefont {W.~G.}\ \bibnamefont
  {Stirling}}, \bibinfo {author} {\bibfnamefont {R.}~\bibnamefont {Scherm}},
  \bibinfo {author} {\bibfnamefont {A.}~\bibnamefont {Stunault}}, \bibinfo
  {author} {\bibfnamefont {B.}~\bibnamefont {Fak}}, \bibinfo {author}
  {\bibfnamefont {H.}~\bibnamefont {Godfrin}}, \ and\ \bibinfo {author}
  {\bibfnamefont {A.~J.}\ \bibnamefont {Dianoux}},\ }\href@noop {} {\bibfield
  {journal} {\bibinfo  {journal} {J. Phys.: Condens. Matter}\ }\textbf
  {\bibinfo {volume} {6}},\ \bibinfo {pages} {821} (\bibinfo {year}
  {1994})}\BibitemShut {NoStop}%
\bibitem [{coh()}]{coherent}%
  \BibitemOpen
  \href@noop {} {}\bibinfo {note} {We focus our presentation on the coherent
  part of the dynamic structure factor, because it is that which is physically
  more interesting, and also more challenging to recover. In general, the
  incoherent part of the spectrum yielded by our approach is in broad
  quantitative agreement with experiment.}\BibitemShut {Stop}%
\bibitem [{\citenamefont {Pearce}\ and\ \citenamefont {Glyde}(2005)}]{pearce}%
  \BibitemOpen
  \bibfield  {author} {\bibinfo {author} {\bibfnamefont {J.~V.}\ \bibnamefont
  {Pearce}}\ and\ \bibinfo {author} {\bibfnamefont {H.~R.}\ \bibnamefont
  {Glyde}},\ }\href@noop {} {\bibfield  {journal} {\bibinfo  {journal} {J. Low
  Temp. Phys.}\ }\textbf {\bibinfo {volume} {138}},\ \bibinfo {pages} {37}
  (\bibinfo {year} {2005})}\BibitemShut {NoStop}%
\bibitem [{pri()}]{private_gift}%
  \BibitemOpen
  \href@noop {} {}\bibinfo {note} {Data from Ref. \cite{gift} were supplied by
  D. E. Galli, private communication. Data from Ref. \cite{ferre} were read off
  Figs. 4 and 7 therein.}\BibitemShut {Stop}%
\bibitem [{\citenamefont {Donnelly}\ \emph {et~al.}(1981)\citenamefont
  {Donnelly}, \citenamefont {Donnelly},\ and\ \citenamefont
  {Hills}}]{donnelly}%
  \BibitemOpen
  \bibfield  {author} {\bibinfo {author} {\bibfnamefont {R.~J.}\ \bibnamefont
  {Donnelly}}, \bibinfo {author} {\bibfnamefont {J.~A.}\ \bibnamefont
  {Donnelly}}, \ and\ \bibinfo {author} {\bibfnamefont {R.~N.}\ \bibnamefont
  {Hills}},\ }\href@noop {} {\bibfield  {journal} {\bibinfo  {journal} {J. Low
  Temp. Phys.}\ }\textbf {\bibinfo {volume} {44}},\ \bibinfo {pages} {471}
  (\bibinfo {year} {1981})}\BibitemShut {NoStop}%
\bibitem [{\citenamefont {Jiang}\ \emph {et~al.}(2001)\citenamefont {Jiang},
  \citenamefont {Jiang},\ and\ \citenamefont {Voth}}]{jiang}%
  \BibitemOpen
  \bibfield  {author} {\bibinfo {author} {\bibfnamefont {S.}~\bibnamefont
  {Jiang}}, \bibinfo {author} {\bibfnamefont {S.}~\bibnamefont {Jiang}}, \ and\
  \bibinfo {author} {\bibfnamefont {G.~A.}\ \bibnamefont {Voth}},\ }\href@noop
  {} {\bibfield  {journal} {\bibinfo  {journal} {J. Chem. Phys.}\ }\textbf
  {\bibinfo {volume} {115}},\ \bibinfo {pages} {7832} (\bibinfo {year}
  {2001})}\BibitemShut {NoStop}%
\end{thebibliography}%

\end{document}